\documentclass[12pt]{article}
  \usepackage{amsfonts}
  \usepackage{amsmath}
\usepackage{amssymb}
\usepackage{amscd}
\usepackage[dvips]{graphicx}
\begin{document}
~~
\bigskip
\bigskip
\begin{center}
\section*
{\Large {\bf{{{Canonical and Lie-algebraic twist deformations of
$\kappa$-Poincare and contractions to $\kappa$-Galilei algebras}}}}}
\end{center}
\bigskip
\bigskip
\bigskip
\begin{center}
{{\large ${\rm {Marcin\;Daszkiewicz}}$}}
\end{center}
\bigskip
\begin{center}
{ {{{Institute of Theoretical Physics\\ University of Wroc{\l}aw pl.
Maxa Borna 9, 50-206 Wroc{\l}aw, Poland\\ e-mail:
marcin@ift.uni.wroc.pl}}}}
\end{center}
\bigskip
\bigskip
\bigskip
\bigskip
\bigskip
\bigskip
\bigskip
\bigskip
\begin{abstract}
~~We propose  canonical  and Lie-algebraic twist deformations of
$\kappa$-deformed Poincare Hopf algebra which leads to the
generalized $\kappa$-Minkowski  space-time relations. The
corresponding deformed $\kappa$-Poincare quantum groups are also
calculated. Finally, we perform the nonrelativistic contraction
limit to the corresponding twisted Galilean algebras and dual
Galilean quantum groups.
\end{abstract}
\bigskip
\bigskip
\bigskip
\bigskip
\eject
\section{{{Introduction}}}

Recently, it has been suggested that the classical Poincare
invariance should be treated as an approximate symmetry in
ultra-high energy regime and the relativistic space-time symmetries
on Planck scale is deformed \cite{1a}-\cite{1d}. Besides, there are
also arguments based on  quantum gravity \cite{2}, \cite{2a} and
string theory models \cite{1s}, \cite{2s} which suggest that
space-time at Planck length is quantum, i.e. it should be
noncommutative. The simplest choice of the noncommutative space-time
is the following
\begin{equation}
[\;{\hat x}_{\mu},{\hat x}_{\nu}\;] = i\theta_{\mu\nu} +
i\theta_{\mu\nu}^{\rho}{\hat
x}_{\rho}\;\;\;;\;\;\;\theta_{\mu\nu},\;
\theta_{\mu\nu}^{\rho}\;-\;{\rm const}\;. \label{noncomm}
\end{equation}

The first, simplest kind of noncommutativity ($\theta_{\mu\nu} \ne
0,\;\theta_{\mu\nu}^{\rho} = 0$ in  formula (\ref{noncomm})) was
 investigated in the Hopf-algebraic framework in
\cite{3e}-\cite{3d}. It corresponds to the well-known canonical
(soft) deformation of Poincare Hopf algebra obtained by twist
procedure \cite{twist}. The second type of space-time deformation
$(\theta_{\mu\nu} = 0,\;\theta_{\mu\nu}^{\rho} \ne 0)$ is directly
associated with another modification of classical relativistic
symmetries - the $\kappa$-deformed Poincare Hopf algebra \cite{4a},
\cite{4b}, which is an example of the Lie-algebraic kind of
space-time noncommutativity.

In almost all  considerations both modifications of Minkowski space
- Lie- and soft-type - are considered separately. Here we  ask about
such a deformation of relativistic space-time symmetry, when both
noncommutativities will appear together, i.e. for which in the
formula (\ref{noncomm}) both coefficients $\theta_{\mu\nu}$ and
$\theta_{\mu\nu}^{\rho}$ are different from  zero.

The results of Zakrzewski's  (\cite{zakrzewski}, \cite{5})  indicate
how to look for such a generalized Hopf Poincare structure. The
classical r-matrix related to such a modification of space-time
symmetries should be a sum of r-matrices for $\kappa$-Poincare group
and the one describing canonical twist. Besides, this extended
r-operator should solve the modified Yang-Baxter equation the same
as in the case of $\kappa$-deformed Poincare symmetry. In this way,
one can see that the explicit form of a proper twist factor allows
us to derive a deformation of new quantum group - canonically
twisted $\kappa$-Poincare Hopf algebra. Moreover, its dual partner
can be calculated by a canonical quantization scheme of
corresponding extended   Poisson-Lie structure \cite{poisson}.

It should be mentioned, that the above algorithm can be generalized
to two other twist deformations of $\kappa$-Poincare algebra -
Lie-type \cite{lukwor}
(see also \cite{lie}) and quadratic-one \cite{lukwor}. 
First of them leads to a Lie-algebraic noncommutativity of Minkowski
space, and it introduces in natural way a second (apart of $\kappa$)
mass-like parameter ${\hat \kappa}$. In the case of quadratic
extension of $\kappa$-Poincare algebra the deformation parameter is
dimensionless.

In this article we consider both the canonical and Lie-algebraic
twist deformations of $\kappa$-Poincare symmetry. In second Section
we recall necessary facts concerning the $\kappa$-deformed Poincare
algebra and its dual quantum group. The canonical and Lie-algebraic
 deformations of $\kappa$-Poincare algebras and
$\kappa$-Minkowski space-times are presented in Section 3 and  4,
respectively.  In Section 5 we find canonically and
Lie-algebraically deformed $\kappa$-Poincare dual groups. Finally,
 the nonrelativistic contraction limits (\cite{con1}-\cite{con3}) to the  twisted Galilean
algebras and dual  quantum groups \cite{gal1}, \cite{con1} are
performed in Section 6. The results are briefly discussed and
summarized in the last Section.

\section{{{$\kappa$-Poincare deformation - short review}}}

\subsection{{{$\kappa$-deformed Poincare algebra}}}

The $\kappa$-deformed Poincare algebra $\,{\mathcal
U}_{\kappa}({\mathcal P})$ is the associative and coassociative Hopf
structure with generators $M_{\mu\nu}$ and $P_{\mu}$ satisfying the
following relations \cite{majid} 
($\eta_{\mu\nu} = (-,+,+,+)$)
\begin{eqnarray}
&&[\;M^{\mu\nu},M^{\lambda\sigma}\;] = i\left(
\eta^{\mu\sigma}M_{\nu\lambda} - \eta^{\nu\sigma}M_{\mu\lambda} +
\eta^{\nu\lambda}M_{\mu\sigma} - \eta^{\mu\lambda}M_{\nu\sigma}
\right)\;,\label{a1}\\
&&~~\vspace{0.3cm} \nonumber \\
&&[\;M^{ij},P_k\;] = i\left(\delta^{j}_{~k}P_i - \delta^i_{~k}P_j\right) \;,\label{a2}\\
&&~~\vspace{0.3cm} \nonumber \\
&&[\;M^{i0},P_j\;] = i\delta^i_{~j}
\left[\frac{\kappa}{2}\left(1-{\rm e}^{-\frac{2P_0}{\kappa}}\right)
+ \frac{1}{2\kappa}\vec{P}^2
\right] - \frac{i}{\kappa}P^iP_j \;, \label{a3}\\
&&~~\vspace{0.3cm} \nonumber \\
&&[\;M^{ij},P_{0}\;] = 0 \;\;,\;\;\;\;[\;M^{i0},P_{0}\;] =
iP_i\;\;\;,\;\;\; [\;P_{\mu},P_{\nu}\;] = 0\;,\label{a5}
\end{eqnarray}
with the coproducts, antipodes and counits defined by
\begin{eqnarray}
&&\Delta_{\kappa}(M^{ij}) = M^{ij}\otimes 1 + 1\otimes
M^{ij}\;,\label{c1}\\
&&~~\vspace{0.3cm} \nonumber \\
&&\Delta_{\kappa}(M^{i0}) = M^{i0}\otimes{\rm
e}^{-\frac{P_0}{\kappa}} + 1\otimes M^{i0} -\frac{1}{\kappa}
M^{ij} \otimes P_j\;,\label{c2}\\
&&~~\vspace{0.3cm} \nonumber \\
&&\Delta_{\kappa}(P_0) = P_0\otimes 1 + 1\otimes P_0\;\;\;,\;\;\;
\Delta_{\kappa}(P_i) = P_i\otimes {\rm e}^{-\frac{P_0}{\kappa}} +
1\otimes P_i\;,\label{c4}
\end{eqnarray}
\begin{eqnarray}
&&S_{\kappa}(M^{ij}) = -M^{ij}\;\;,\;\;S_{\kappa}(M^{i0}) = -
\left(M^{i0} + \frac{1}{\kappa}M^{ij}P_j \right){\rm e}^{\frac{P_0}{\kappa}}\;,\label{anti1}\\
&&~~\vspace{0.3cm} \nonumber \\
&& S_{\kappa}(P_i) = -P_i{\rm
e}^{\frac{P_0}{\kappa}}\;\;,\;\;S_{\kappa}(P_0) = -P_0\;\;,\;\;
\epsilon (P_\mu) = \epsilon(M^{\mu\nu}) = 0\;. \label{anti2}
\end{eqnarray}
The $\kappa$-deformed mass Casimir looks as follows
\begin{equation}
C_{\kappa} = \left(2\kappa {\rm sinh}\left(\frac{P_0}{\kappa}\right)
\right)^2 -\overrightarrow{P}^2{\rm e}^{\frac{P_0}{\kappa}}\;.
\label{casimir}
\end{equation}
We see that in $\,{\mathcal U}_{\kappa}({\mathcal P})$ one can
distinguish the following two Hopf subalgebras:  non-deformed
$O(3)$-rotation algebra and Abelian fourmomentum algebra. For
$\kappa \to \infty$ the deformation disappears and we get the
classical Poincare Hopf algebra $\,{\mathcal U}_0({\mathcal P})$.

It is well-known that the classical r-matrix corresponding to the
above Hopf structure has the form \cite{zakrzewski},
\cite{zakrzewskigr}, \cite{maslanka}
\begin{equation}
r_{\kappa} = \frac{1}{\kappa}M_{0\mu} \wedge P^{\nu} =
r^{\mu\nu;\alpha}_{\kappa} M_{\mu\nu} \wedge P_{\alpha}\;\;\;;\;\;\;
r^{\mu\nu;\alpha}_{\kappa} =
\frac{1}{2\kappa}\left(\delta^{\mu}_{~0}\eta^{\nu\alpha} -
\delta^{\nu}_{~0}\eta^{\mu\alpha} \right)\;, \label{rkappa1}
\end{equation}
with $a \wedge b = a\otimes b - b\otimes a$. One can check that the
matrix (\ref{rkappa1}) with itself satisfies a modified Yang-Baxter
equation (MYBE)
\begin{equation}
[[\;r_{\kappa},r_{\kappa}\;]]:= [\;r_{\kappa12},r_{\kappa13} +
r_{\kappa23}\;] + [\;r_{\kappa13}, r_{\kappa23}\;] =
\frac{1}{\kappa^2} M_{\mu\nu} \wedge P^{\mu} \wedge P^{\nu}\;,
\label{mybe}
\end{equation}
where used in the above formula symbol $[[\;\cdot,\cdot\;]]$ denotes
Schouten bracket while $r_{\kappa 12} = \frac{1}{\kappa}M_{i0}\wedge
P_i\wedge 1$, $r_{\kappa 13} =  \frac{1}{\kappa}M_{i0}\wedge 1
\wedge P_i$ and $r_{\kappa 23} =  \frac{1}{\kappa} 1\wedge
M_{i0}\wedge P_i$.

\subsection{{{$\kappa$-deformed Poincare group}}}

The classical r-matrix (\ref{rkappa1}) defines  Poisson-Lie
structure \cite{poisson}. Its standard quantization procedure leads
to a dual form of the Hopf algebra (\ref{a1})-(\ref{anti2}) - the
$\kappa$-deformed Poincare group ${\mathcal P}_{\kappa}$
\cite{zakrzewskigr}, \cite{maslanka}. It is defined by the following

a) algebraic relationes
\begin{eqnarray}
&&[\,\Lambda^{\alpha}_{~\beta},a^{\rho}\,] = -\frac{i}{\kappa}
((\Lambda^{\alpha}_{~0} -
\delta^{\alpha}_{~0})\Lambda^{\rho}_{~\beta} +
\eta^{\alpha\rho}(\Lambda_{0\beta} - \eta_{0\beta}))\;,\label{gkappa1} \\
&&~~\vspace{0.3cm} \nonumber \\
&&~~[\,a^{\rho},a^{\sigma}\,] = -\frac{i}{\kappa}
(\delta^{\sigma}_{~0}a^{\rho} -
\delta^{\rho}_{~0}a^{\sigma})\;\;\;,\;\;\;
[\,\Lambda^{\alpha}_{~\beta},\Lambda^{\delta}_{~\rho}\,] =
0\;,\label{gkappa3}
\end{eqnarray}

b) coproducts
\begin{eqnarray}
&&~~~~\Delta(\Lambda^{\mu}_{~\nu}) = \Lambda^{\mu}_{~\alpha}\otimes
\Lambda^{\alpha}_{~\nu}\;\;\;,\;\;\;\Delta(a^{\mu}) =
\Lambda^{\mu}_{~\nu}\otimes a^{\nu} + a^{\mu} \otimes
1\;,\label{gkappa5}
\end{eqnarray}

c) antipodes and counits
\begin{eqnarray}
S(\Lambda^{\mu}_{~\nu}) = \Lambda^{\mu}_{~\nu} \;\;\;,\;\;\;
S(a^{\mu}) = -\Lambda^{\mu}_{~\nu} a^{\mu}\;\;\;,\;\;\;
\epsilon(\Lambda^{\mu}_{~\nu}) = \delta^{\mu}_{~\nu} \;\;\;,\;\;\;
\epsilon(a^{\mu}) = 0\;.\label{gkappa6}
\end{eqnarray}
The used above generators $\Lambda^{\mu}_{~\nu}$ are dual to
$M^{\mu\nu}$ - Lorentz rotation generators
\begin{eqnarray}
< \Lambda^{\mu}_{~\nu},M^{\alpha\beta} > \,= \left(\eta^{\alpha\mu}
\delta^{\beta}_{~\nu} - \eta^{\beta\mu} \delta^{\alpha}_{~\nu}
\right)\;,
\end{eqnarray}
while $a^{\mu}$ are dual to $P_{\mu}$ (translations)
\begin{eqnarray}
< a^{\mu},P_{\nu} > \,= \delta^{\mu}_{~\nu}\;.
\end{eqnarray}
It should be noted that the relations
(\ref{gkappa5})-(\ref{gkappa6}) remain undeformed as for the
classical Poincare group ${\mathcal P}$.

\subsection{{{$\kappa$-deformed Minkowski space}}}

It is well-known (see e.g. \cite{dlw14b}) that the deformed
Minkowski space can be introduced as the quantum representation
space (a Hopf module) for quantum Poincare algebra, equipped with a
proper defined $\star$-multiplication of two arbitrary function.
Such a $\star$-product should be consistent with the action of
deformed symmetry generators satisfying suitably deformed Leibnitz
(coproduct) rules. In the case of $\kappa$-deformation
the 
 $\star_{\kappa}$-multiplication looks as follows (see
\cite{starmulti}, \cite{lit12a}  and references therein)
\begin{equation}
f(x)\star_{\kappa} g(x) = \omega \left({\mathcal
O}_{\kappa}(x_\mu,\partial^{\mu}) (f(x)\otimes g(x))\right) \;,
\label{kappastar}
\end{equation}
where $\,\omega(f(x)\otimes g(x)) = f(x)g(x)$ and the
$\star_{\kappa}$-differential operator is given by
\begin{equation}
{\mathcal O}_{\kappa}(x_\mu,\partial^{\mu}):= {\rm
exp}(ix_{\mu}\gamma^{\mu}(\partial^{\nu})) \;, \label{kappastarop}
\end{equation}
with
\begin{equation}
\gamma^{\mu}(\partial^{\nu}):= c^{\mu}_{~\rho\tau}
\partial^{\rho}\otimes \partial^{\tau}
+\frac{1}{12}c^{\mu}_{~\rho\tau}c^{\rho}_{~\lambda\nu}
(\partial^{\tau}\partial^{\lambda}\otimes \partial^{\nu} +
\partial^{\nu}\otimes \partial^{\tau}\partial^{\lambda})+
\cdots \;; \label{gamma}
\end{equation}
\begin{equation}
c^{i}_{~0i} = -c^{i}_{~i0} = -\frac{1}{2\kappa}\;\;\; {\rm
other}\;\;\; c^{\mu}_{~\rho\tau} = 0\;. \label{c}
\end{equation}
Using the formula (\ref{kappastar})  in the case
$f(x)=x_\mu,\;g(x)=x_\nu$ we see that the $\kappa$-deformed
Minkowski space-time  takes the form
\begin{eqnarray}
&&~~~~[\,x_{i},x_{0}\,]_{\star_{\kappa}} =
x_{i}{\star_{\kappa}}x_{0} - x_{0}{\star_{\kappa}}x_{i} =
\frac{i}{\kappa}
x_{i}\;\;\;,\;\;\;[\,x_{i},x_{j}\,]_{\star_{\kappa}} = 0\;,
\end{eqnarray}
and in the $\kappa \to \infty$ limit it becomes classical.


\section{{{Canonical  twist deformation of
$\kappa$-Poincare algebra}}}

\subsection{{{Extended classical r-matrix}}}

Let us consider the following extension of classical r-matrix
(\ref{rkappa1})
\begin{equation}
r = r_{\kappa} + r_{{\hat \kappa}} + r_{\xi}\;, \label{rmatrix}
\end{equation}
with
\begin{equation}
r_{{\hat \kappa}} = \frac{1}{2{\hat \kappa}}M_{12} \wedge P_{0}\;,
\label{rmatrix1}
\end{equation}
and
\begin{equation}
r_{\xi} = \frac{\xi}{2} P_{3} \wedge P_{0}\;,\label{rmatrix2}
\end{equation}
where the formulas (\ref{rmatrix1}) and (\ref{rmatrix2}) describe
Lie-algebraic and canonical twist deformations of $\kappa$-Poincare
algebra, respectively. Due to the commutation relations
$[\;P_{\mu},P_{\nu}\;] = [\;M_{12},P_3\;] = 0$ (see (\ref{a2}) and
(\ref{a5})) we can see that both matrices $r_{{\hat \kappa}}$ and
$r_{\xi}$ satisfy the classical Yang-Baxter equation (CYBE)
\begin{equation}
[[\;r_{{\hat \kappa}},r_{{\hat \kappa}}\;]] =
[[\;r_{\xi},r_{\xi}\;]] = 0\;;  \label{cybe}
\end{equation}
the mixed Schouten brackets vanish as well
\begin{equation}
[[\;r_{{\hat \kappa}},r_{\xi}\;]] = [[\;r_{\xi},r_{{\hat
\kappa}}\;]] = 0\;. \label{mcybe}
\end{equation}
By explicit calculation one can  check that 
\begin{equation}
~~~~[[\;r_{{\kappa}},r_{\cdot}\;]] = [[\;r_{\cdot},r_{{\kappa}}\;]]
= 0\;\;\;;\;\;\;r_{\cdot} = r_{{\hat \kappa}},\;r_{\xi}\;,
\label{kon}
\end{equation}
which together with the formulas (\ref{cybe}) and (\ref{mcybe})
means that the extended r-matrix (\ref{rmatrix}) satisfies the
modified Yang-Baxter equation (\ref{mybe})
\begin{equation}
[[\;r,r\;]] = \frac{1}{\kappa^2} M_{\mu\nu} \wedge P^{\mu} \wedge
P^{\nu}\;.  \label{rmybe}
\end{equation}

\subsection{{{Canonical deformation of $\kappa$-Poincare algebra}}}

In accordance with (\ref{kon}) one can  consider the canonical $(r =
r_{\kappa} + r_{\xi})$ deformation of enveloping $\kappa$-Poincare
algebra $\,{\mathcal U}_{\kappa}({\mathcal P})$. As already
mentioned in Introduction we can get such a modification of
space-time relativistic symmetry by a proper ($\kappa$-deformed)
twisting procedure.

First of all, let us introduce an element ${\mathcal F}_{\xi} \in
\,{\mathcal U}_{\kappa}({\mathcal P}) \otimes \,{\mathcal
U}_{\kappa}({\mathcal P})$ with the following linear term in series
expansion with respect to the deformation parameter $\xi$
\begin{equation}
{\cal F}_{\xi} = 1 + ir_{\xi}^{(1)} \otimes r_{\xi}^{(2)} +
\cdots\;\;;\;\;\;r_{\xi} = r_{\xi}^{(1)} \otimes r_{\xi}^{(2)}\;.
\label{expansion}
\end{equation}
Next, we define Drinfeld twist factor as the function
(\ref{expansion}) satisfying so-called $\kappa$-deformed cocycle
condition \cite{drin}
\begin{equation}
{\cal F}_{\xi12} \cdot(\Delta_{\kappa} \otimes 1) ~{\cal F}_{\xi} =
{\cal F}_{\xi23} \cdot(1\otimes \Delta_{\kappa}) ~{\cal F}_{\xi}\;,
\label{cocycle}
\end{equation}
and the normalization condition
\begin{equation}
(\epsilon \otimes 1)~{\cal F}_{\xi} = (1 \otimes \epsilon)~{\cal
F}_{\xi} = 1\;, \label{normalization}
\end{equation}
with ${\cal F}_{\xi12} = {\cal F}_{\xi}\otimes 1$ and ${\cal
F}_{\xi23} = 1 \otimes {\cal F}_{\xi}$. The solution of above
equations has been found in \cite{5} 
and it looks as follows
\begin{equation}
{\cal F}_{\xi,\kappa} = {\rm exp}\left(i{\kappa \frac{\xi}{2} P_3
\otimes\left({\rm e}^{-\frac{P_0}{\kappa}} -1 \right)}\right)\;.
\label{solution}
\end{equation}
One can easily see that in the limit $\xi \to 0$ factor ${\cal
F}_{\xi,\kappa}$ goes to the unit operator
\begin{equation}
\lim_{{\xi \to 0}}\,{\cal F}_{\xi,\kappa} = 1\;, \label{limit}
\end{equation}
while in the case   $\kappa \to \infty$  we get a standard
canonical-twist element for the classical Poincare Hopf algebra
$\,{\mathcal U}_0({\mathcal P})$
\begin{equation}
{\cal F}_{\xi,\infty} = {\rm e}^{-i\frac{\xi}{2} P_3\otimes P_0}\;.
\label{limkappa}
\end{equation}

It is well-known that twist ${\cal F}_{\xi,\kappa}$ does not modify
the algebraic part of $\kappa$-Poincare algebra
(\ref{a1})-(\ref{a5}) and counits, but it  changes the coproducts
(\ref{c1})-(\ref{c4}) and antipodes (\ref{anti1}), (\ref{anti2})
according to
\begin{eqnarray}
&&~~~~~~~~\Delta_{{\cal F}_{\xi,\kappa}}(a) = {\cal
F}_{\xi,\kappa}\;\Delta_{{\kappa}}(a)\;{\cal F}_{\xi,\kappa}^{-1}\;,
\label{d1} \\
&~~&  \cr &&S_{{\cal F}_{\xi,\kappa}}(a) =
u(\kappa,\xi)~S_{\kappa}(a)~u^{-1}(\kappa,\xi)\;,~~~~~~~~
\end{eqnarray}
where $u(\kappa,\xi) = \sum f_{(1)}S_{\kappa}(f_{(2)})$, and where
we use Sweedler's notation ${\cal F}_{\xi,\kappa} = \sum f_{(1)}
\otimes f_{(2)}$. Hence, using the formula
\begin{equation}
u(\kappa,\xi) = {\rm exp}\left(i{\kappa \xi P_3 \left(
\exp({P_0}/{\kappa})-1 \right)}\right)\;, \label{u}
\end{equation}
we obtain
\begin{eqnarray}
\Delta_{{\cal F}_{\xi,\kappa}}(P_0) &=&  P_0\otimes 1 + 1\otimes
P_0\;\;\;,\;\;\;
\Delta_{{\cal F}_{\xi,\kappa}}(P_i) = P_i\otimes {\rm e}^{-\frac{P_0}{\kappa}} + 1\otimes P_i\;,\label{modcopro1}\\
&~~&  \cr \Delta_{{\cal F}_{\xi,\kappa}}(M^{ij}) &=&
\Delta_{\kappa}(M^{ij}) + \kappa \frac{\xi}{2} \left(
\delta^j_{~3}P_i - \delta^i_{~3}P_j\right)
\otimes \left({\rm e}^{-\frac{P_0}{\kappa}} -1 \right)\;, \\
 &~~&  \cr
\Delta_{{\cal F}_{\xi,\kappa}}(M^{i0}) &=& \Delta_{\kappa}(M^{i0}) -
\frac{\xi}{2} P_3 \otimes P_i{\rm e}^{-\frac{P_0}{\kappa}} 
+\\&~~&  \cr &+&{\frac{\xi}{2}}\left( \delta^{i}_{~3}
\left[\frac{\kappa}{2}\left(1-{\rm e}^{-\frac{2P_0}{\kappa}}\right)
+ \frac{1}{2\kappa}\vec{P}^2
\right] + \frac{1}{\kappa}P_iP_3 \right) \otimes \\
&~~&  \cr &\otimes& \kappa\left({\rm e}^{-\frac{P_0}{\kappa}} -1
\right){\rm e}^{-\frac{P_0}{\kappa}}+
\\
&~~&  \cr
 &+& {\frac{\xi}{2}}\left( \delta^j_{~3}P_i - \delta^i_{~3}P_j\right)
\otimes P_j\left({\rm e}^{-\frac{P_0}{\kappa}} -1 \right)\;,
\label{modcopro2}
\end{eqnarray}
and
\begin{eqnarray}
S_{{\cal F}_{\xi,\kappa}}(P_0) &=& S_{\kappa}(P_0) =
-P_0\;\;\;,\;\;\;S_{{\cal F}_{\xi,\kappa}}(P_i) =
S_{\kappa}(P_i) = -P_i{\rm e}^{\frac{P_0}{\kappa}}\;,\label{modanti1}\\
&~~&  \cr S_{{\cal F}_{\xi,\kappa}}(M^{ij}) &=&  S_{\kappa}(M^{ij})-
\kappa\xi \left(\delta^{j}_{~3}P_i -
\delta^i_{~3}P_j\right)\cdot\left(\exp({P_0}/{\kappa})-1
\right) \;,
\\
&~~&  \cr S_{{\cal F}_{\xi,\kappa}}(M^{i0}) &=& S_{\kappa}(M^{i0}) -
\xi\left(\delta^{j}_{~3}P_i - \delta^i_{~3}P_j\right) P_j\cdot{\rm
e}^{\frac{P_0}{\kappa}} \cdot \\ &~~&  \cr &\cdot&
\left(\exp({P_0}/{\kappa})-1 \right)-\xi P_3P_i 
{\rm e}^{\frac{2P_0}{\kappa}}+\\
&~~&  \cr &-&\kappa\xi\left( \delta^i_{~3}
\left[\frac{\kappa}{2}\left(1-{\rm e}^{-\frac{2P_0}{\kappa}}\right)
+ \frac{1}{2\kappa}\vec{P}^2 \right] - \frac{1}{\kappa} P_iP_3\right)\cdot\\
&~~&  \cr
 &\cdot&{\rm
e}^{\frac{P_0}{\kappa}}\cdot \left(\exp({P_0}/{\kappa})-1 \right)
\;. \label{modanti2}
\end{eqnarray}
The algebraic relations (\ref{a1})-(\ref{a5}) together with
coproducts (\ref{modcopro1})-(\ref{modcopro2}), antipodes
(\ref{modanti1})-(\ref{modanti2}) and classical counits
(\ref{anti2}) define the canonical twist deformation of
$\kappa$-Poincare algebra $\,{\mathcal U}_{{\xi,\kappa}}({\mathcal
P})$. We see, that for $\xi \to 0$ one gets the $\kappa$-Poincare
algebra $\,{\mathcal U}_{{\kappa}}({\mathcal P})$, which  is in
accordance with the formula (\ref{limit}), i.e. there is no twist
transformation in such a case. For parameter $\kappa \to \infty$,
the algebra $\,{\mathcal U}_{{\xi,\kappa}}({\mathcal P})$ passes
into well-known  ($\theta^{\mu\nu}$)-Poincare Hopf structure
\cite{3a},
and this time, it agrees with
the form of twist factor (\ref{limkappa}).


\subsection{{{Canonical extension of $\kappa$-Minkowski space}}}

Let us now find a noncommutative Minkowski space corresponding to
the canonical deformation of $\kappa$-Poincare. As it was mentioned
in the first section, our space-time can be defined as a quantum
representation space for the extended quantum Poincare algebra
$\,{\mathcal U}_{{\xi,\kappa}}({\mathcal P})$, equipped with a
proper deformed $\star$-multiplication. We define our
$\star$-product for arbitrary two functions depending on space-time
coordinates as follows
\begin{equation}
f(x)\star_{{\kappa},\xi} g(x) = \omega \left({\mathcal
O}_{{\kappa},\xi}(x_\mu,\partial^{\mu}) (f(x)\otimes g(x))\right)
\;, \label{extstar}
\end{equation}
where the $\star$-operator ${\mathcal
O}_{\star_{{\kappa},\xi}}(x_\mu,\partial^{\mu})$ is given by the
superposition of two $\star$-operators: for the $\kappa$-deformed
r-matrix $r_{\kappa}$ (see (\ref{kappastarop})), and  for the
canonical deformed matrix $r_{\xi}$ (see twist factor
(\ref{solution})) \cite{dlw14b}
\begin{equation}
{\mathcal O}_{\xi}(x_\mu,\partial^{\mu}):={\cal
F}^{-1}_{\xi,\kappa}(x_\mu,\partial^{\mu}) = {\rm
exp}\left(-i{\kappa \frac{\xi}{2} \partial^3 \otimes\left({\rm
e}^{-\frac{\partial^0}{\kappa}} -1 \right)}\right)\;. \label{ext2}
\end{equation}
Consequently, our operator takes the form
\begin{equation}
{\mathcal O}_{{\kappa},\xi}(x_\mu,\partial^{\mu}):={\mathcal
O}_{\xi}(x_\mu,\partial^{\mu}) \circ {\mathcal
O}_{{\kappa}}(x_\mu,\partial^{\mu})
 \;,
\label{extstarop}
\end{equation}
and we obtain the following  commutation relations
\begin{eqnarray}
[\,x_{i},x_{0}\,]_{\star_{{\kappa},\xi}} = \frac{i}{\kappa} x_{i} +
i\frac{\xi}{2} \delta^3_{~i}\;\;\;,\;\;\;
[\,x_{i},x_{j}\,]_{\star_{{\kappa},\xi}} =
0\;.\label{dst2}
\end{eqnarray}
The relations  (\ref{dst2})  define the canonically extended
$\kappa$-Minkowski space-time ${\mathcal M}_{{\kappa},\xi}$. We see
that the soft deformation of $\kappa$-Poincare algebra introduces
two kinds of  noncommutativity: Lie-type associated with parameter
${\kappa}$, and  canonical type - corresponding to  parameter $\xi$.
Of course, for $\xi \to 0$ one gets the $\kappa$-deformed Minkowski
space-time ${\mathcal M}_{\kappa}$, while in the ${\kappa} \to
\infty$ limit we obtain  well-known $\theta^{\mu\nu}$-deformed
Minkowski space ${\mathcal M}_{\theta}$ (see e.g. \cite{3a}).

\section{{{Lie-algebraic twist deformation of
$\kappa$-Poincare algebra}}}

\subsection{{{Deformation of algebra}}}

In the case of Lie-algebraic deformation
$(r=r_{\kappa}+r_{\hat{\kappa}})$ the  twist factor  has been found
in \cite{5}. Here we consider its antisymmetric form
\begin{equation}
{\cal F}_{\hat{\kappa},\kappa} = {\rm
exp}\left(\frac{i}{2\hat{\kappa}}M_{12} \wedge P_0\right)\;.
\label{solution1e}
\end{equation}
By tedious calculation we get the following coproduct of deformed
$\kappa$-Poincare algebra $\,{\mathcal U}_{{{\hat
\kappa},\kappa}}({\mathcal P})$
\begin{eqnarray}
\Delta_{{\cal F}_{\hat{\kappa},\kappa}}(P_0) &=&  P_0\otimes 1 +
1\otimes P_0\;\;\;,\;\;\;
\Delta_{{\cal F}_{\hat{\kappa},\kappa}}(P_3)  = P_3\otimes {\rm e}^{-\frac{P_0}{\kappa}} + 1\otimes P_3\;,\label{fmodcopro1}\\
&~~&  \cr \Delta_{{\cal F}_{\hat{\kappa},\kappa}}(P_1) &=&
\Delta_{\kappa}(P_1) - \sin\left(\frac{P_0}{2{\hat \kappa}} \right)
\otimes P_2 + P_2\otimes \sin\left(\frac{P_0}{2{\hat
\kappa}}\right){\rm
e}^{-\frac{P_0}{\kappa}}+\\
&~~&  \cr &-&\left[ \cos\left(\frac{P_0}{2{\hat \kappa}}
\right)-1\right]\otimes P_1 - P_1\otimes\left[
\cos\left(\frac{P_0}{2{\hat \kappa}} \right)-1\right]{\rm
e}^{-\frac{P_0}{\kappa}}\;,\\
&~~&  \cr \Delta_{{\cal F}_{\hat{\kappa},\kappa}}(P_2) &=&
\Delta_{\kappa}(P_2) + \sin\left(\frac{P_0}{2{\hat \kappa}} \right)
\otimes P_1 - P_1\otimes \sin\left(\frac{P_0}{2{\hat
\kappa}}\right){\rm
e}^{-\frac{P_0}{\kappa}}+\\
&~~&  \cr &-&\left[ \cos\left(\frac{P_0}{2{\hat \kappa}}
\right)-1\right]\otimes P_2 - P_2\otimes\left[
\cos\left(\frac{P_0}{2{\hat \kappa}} \right)-1\right]{\rm
e}^{-\frac{P_0}{\kappa}}\;,\\
&~~&  \cr \Delta_{{\cal F}_{\hat{\kappa},\kappa}}(M^{ij}) &=&
\Delta_{\kappa}(M^{ij}) - i[\;M^{ij},M^{12}\;]\wedge
\sin\left(\frac{P_0}{2{\hat \kappa}}\right)+\\
&~~&  \cr &~~&~~~~~~~~~~~~~~~-[\;[\;M^{ij},M^{12}\;],M^{12}\;]\perp
\left[ \cos\left(\frac{P_0}{2{\hat \kappa}} \right)-1\right]\;,
\end{eqnarray}
\begin{eqnarray}
&&\Delta_{{\cal F}_{\hat{\kappa},\kappa}}(M^{i0}) =\nonumber\\
&=&\Delta_{\kappa}(M^{i0}) - \frac{1}{2\hat{\kappa}} M^{12} \otimes
P_i + \frac{1}{2\hat{\kappa}} P_i\otimes M^{12}{\rm
e}^{-\frac{P_0}{\kappa}}
 - i[\;M^{i0},M^{12}\;]\otimes
\sin\left(\frac{P_0}{2{\hat \kappa}}\right){\rm
e}^{-\frac{P_0}{\kappa}} \nonumber\\
&+& i\sin\left(\frac{P_0}{2{\hat \kappa}}\right)\otimes
[\;M^{i0},M^{12}\;]- [\;[\;M^{i0},M^{12}\;],M^{12}\;]\otimes\left[
\cos\left(\frac{P_0}{2{\hat \kappa}} \right)-1\right]{\rm
e}^{-\frac{P_0}{\kappa}}\nonumber
\\
&~~&  \cr &-& \left[ \cos\left(\frac{P_0}{2{\hat \kappa}}
\right)-1\right]
\otimes[\;[\;M^{i0},M^{12}\;],M^{12}\;]-\frac{1}{2\hat{\kappa}}
M^{12}\sin\left(\frac{P_0}{2{\hat \kappa}}\right)\otimes \left(
\delta^{1i}P_2 - \delta^{2i}P_1\right) \nonumber
\end{eqnarray}
\begin{eqnarray}
~~~~~~~~~~ &-&\frac{1}{2\hat{\kappa}}\left( \delta^{1i}P_2 -
\delta^{2i}P_1\right)\otimes M^{12}\sin\left(\frac{P_0}{2{\hat
\kappa}}\right) {\rm e}^{-\frac{P_0}{\kappa}}\nonumber~\\
&+&\frac{1}{2\hat{\kappa}}\left( \delta^{1i}P_1 +
\delta^{2i}P_2\right)\otimes M^{12}\left[
\cos\left(\frac{P_0}{2{\hat \kappa}} \right)-1\right]{\rm
e}^{-\frac{P_0}{\kappa}}\label{fmodcopro2}
\end{eqnarray}
\begin{eqnarray}
 &-&\frac{1}{2\hat{\kappa}}M^{12}\left[
\cos\left(\frac{P_0}{2{\hat \kappa}} \right)-1\right]\otimes \left(
\delta^{1i}P_1 +
\delta^{2i}P_2\right)+\frac{i}{{\kappa}}[\;M^{ij},M^{12}\;]\otimes\sin\left(\frac{P_0}{2{\hat
\kappa}}\right)P_j \nonumber\\
 &~~&  \cr
 &+&
\frac{1}{{\kappa}}[\;[\;M^{ij},M^{12}\;],M^{12}\;]\otimes\left[
\cos\left(\frac{P_0}{2{\hat \kappa}} \right)-1\right]P_j
+\frac{1}{{\kappa}}\sin\left(\frac{P_0}{2{\hat
\kappa}}\right)M^{ij}\otimes\left( \delta^{1j}P_2 -
\delta^{2j}P_1\right)\nonumber
\end{eqnarray}
\begin{eqnarray}
~~~~~~~~\, &+& \frac{1}{{\kappa}}\left[ \cos\left(\frac{P_0}{2{\hat
\kappa}} \right)-1\right]M^{ij}\otimes \left( \delta^{1j}P_1 +
\delta^{2j}P_2\right)\;. \nonumber
\end{eqnarray}
\\
The algebraic sector as well as the antipodes remain
$\kappa$-deformed, i.e. $S_{{\cal F}_{\hat{\kappa},\kappa}}(a) =
S_{\kappa}(a)$ (see (\ref{anti1}), (\ref{anti2})).


\subsection{{{Two-parameter extension of  $\kappa$-Minkowski space}}}

For Lie-algebraic deformation we define the $\star$-operator
 as follows
\begin{equation}
{\mathcal
O}_{{\kappa},\hat{\kappa}}(x_\mu,\partial^{\mu}):={\mathcal
O}_{\hat{\kappa}}(x_\mu,\partial^{\mu})\circ {\mathcal
O}_{{\kappa}}(x_\mu,\partial^{\mu})
 \;;
\label{extstarop11}
\end{equation}
\begin{equation}
{\mathcal O}_{\hat{\kappa}}(x_\mu,\partial^{\mu}):= {\cal
F}^{-1}_{\hat{\kappa},\kappa}(x_\mu,\partial^{\mu}) = {\rm
exp}\left(-\frac{i}{2\hat{\kappa}}(x_{1}\partial^{2} -
x_{2}\partial^{1})\wedge\partial^{0}\right) \;, \label{ext33}
\end{equation}
and our $({\kappa},\hat{\kappa})$-deformed Minkowski space takes the
form
\begin{eqnarray}
[\,x_{i},x_{0}\,]_{\star_{{\kappa},\hat{\kappa}}} = \frac{i}{\kappa}
x_{i} + \frac{i}{{\hat\kappa}}(\delta^{1}_{~i}x_{2} -
\delta^{2}_{~i}x_{1}) \;\;\;,\;\;\;
[\,x_{i},x_{j}\,]_{\star_{{\kappa},\hat{\kappa}}} =
0\;.\label{dst2ff}
\end{eqnarray}
The relations  (\ref{dst2ff})  define the Lie-algebraic extension of
$\kappa$-Minkowski space-time ${\mathcal
M}_{{\kappa},\hat{\kappa}}$. We see that  above
 deformation of $\kappa$-Poincare algebra introduces Lie-algebraic type of
space-time noncommutativity corresponding to  both parameters
${\kappa}$ and ${\hat \kappa}$. It should be also noted that in the
${\hat \kappa} \to \infty$ limit we get the $\kappa$-deformed
Minkowski space-time ${\mathcal M}_{\kappa}$, while for ${\kappa}
\to \infty$ we obtain the Minkowski space for  Lie-twisted 
Poincare algebra ${\mathcal M}_{\hat{\kappa}}$ \cite{lie}.

\section{Canonical and Lie-algebraic  twist deformation of
$\kappa$-Poincare group}

In accordance with the equation (\ref{rmybe}) one can define the
 corresponding to the matrix (\ref{rmatrix}) Poisson-Lie
structure as follows \cite{poisson}
\begin{equation}
\{\,f,g\,\} = 2r^{AB} \left( X_{A}^RfX_{B}^Rg -
X_{A}^LfX_{B}^Lg\right)\;. \label{poissonn}
\end{equation}
The symbols $ X_{A}^R$, $ X_{A}^L$ denote the right- and
left-invariant vector fields on classical Poincare group ${\mathcal
P}$ given by
\begin{eqnarray}
&&~~~~~~X^{\alpha\beta}_L =
\Lambda^{\mu\alpha}\frac{\partial}{\partial \Lambda^{\mu}_{~\beta}}
- \Lambda^{\mu\beta}\frac{\partial}{\partial
\Lambda^{\mu}_{~\alpha}} \;\;\;,\;\;\; X^{\alpha}_L =
\Lambda^{\mu\alpha}\frac{\partial}{\partial a^{\mu}}
\;, \label{fields} \\
&&~~\vspace{0.3cm} \nonumber \\
&&X^{\alpha\beta}_R =\Lambda^{\beta}_{~\nu}\frac{\partial}{\partial
\Lambda^{\alpha\nu}}
-\Lambda^{\alpha}_{~\nu}\frac{\partial}{\partial \Lambda^{\beta\nu}}
+ a^{\beta}\frac{\partial}{\partial a_{\alpha}} -
a^{\alpha}\frac{\partial}{\partial a_{\beta}} \;\;\;,\;\;\;
X^{\alpha}_R = \frac{\partial}{\partial a_{\alpha}}\;.
\label{fields3}
\end{eqnarray}
If we calculate the Poisson brackets (\ref{poissonn}) with use of
the formulas (\ref{rkappa1}), (\ref{rmatrix1}) and (\ref{rmatrix2}),
in a first step, and if we perform its standard quantization by
replacing $\{\,\cdot,\cdot\,\} \to \frac{1}{i}[\,\cdot,\cdot\,]$, as
a second step, then we obtain the following set of commutation
relations
\begin{eqnarray}
&&[\,\Lambda^{\alpha}_{~\beta},a^{\rho}\,] = -\frac{i}{\kappa}
((\Lambda^{\alpha}_{~0} -
\delta^{\alpha}_{~0})\Lambda^{\rho}_{~\beta} +
\eta^{\alpha\rho}(\Lambda_{0\beta} -
\eta_{0\beta}))\;+~~~~~~~~~~~~~~~~~~~~~~~~\label{gr1}\\
&&~~\vspace{0.3cm} \nonumber \\
&~~& ~~~~~~~~~~~~~~~~+\frac{1}{{\hat\kappa}} (\Lambda^{\rho}_{~0}(
\eta_{2\beta}\Lambda^{\alpha}_{~1} -
\eta_{1\beta}\Lambda^{\alpha}_{~2}) + \delta^{\rho}_{~0}(
\delta^{\alpha}_{~2}\Lambda_{1\beta} -
\delta^{\alpha}_{~1}\Lambda_{2\beta})) \;,
 \end{eqnarray}
\begin{eqnarray}
[\,a^{\rho},a^{\sigma}\,] &=& -\frac{i}{\kappa} (
\delta^{\sigma}_{~0}a^{\rho} - \delta^{\rho}_{~0}a^{\sigma}) +
\frac{i}{{\hat\kappa}}\delta^{\sigma}_{~0}(\delta^{\rho}_{~2}a^{1} -
\delta^{\rho}_{~1}a^{2})\; + \\
&&~~\vspace{0.3cm} \nonumber \\
&+&\frac{i}{{\hat\kappa}}\delta^{\rho}_{~0}(\delta^{\sigma}_{~1}a^{2}
- \delta^{\sigma}_{~2}a^{1}) +i
\frac{\xi}{2}(\delta^{\rho}_{~3}\delta^{\sigma}_{~0} -
\delta^{\rho}_{~0}\delta^{\sigma}_{~3}) + \\
&&~~\vspace{0.3cm} \nonumber \\
&+&i\frac{\xi}{2}(\Lambda^{\rho}_{~0}\Lambda^{\sigma}_{~3} -
\Lambda^{\rho}_{~3}\Lambda^{\sigma}_{~0})\;\;\;,\;\;\;[\,\Lambda^{\alpha}_{~\beta},\Lambda^{\delta}_{~\rho}\,]
= 0\label{gr2}\;.
\end{eqnarray}
Next, if we define the $\ast$-operation in such a way that
$\Lambda^{\mu}_{~\nu}$ and $a^{\mu}$ are selfadjoint elements, we
see that the above relations together with  coproducts
 (\ref{gkappa5}), counits and antipodes
(\ref{gkappa6}) give a Hopf $\ast$-algebra - the
$(\hat{\kappa},\xi)$-{deformed $\kappa$-Poincare group} ${\mathcal
P}_{\kappa,\hat{\kappa},\xi}$. In such a way for $\hat{\kappa} =
\infty$ we get  dual group
to the canonically deformed algebra $\,{\mathcal
U}_{{{\kappa},\xi}}({\mathcal P})$, while for $\xi =0$ we obtain
dual partner
for $\,{\mathcal U}_{{{\kappa},\hat{\kappa}}}({\mathcal P})$.

It should be also noted that  for ${\kappa} \to \infty$,
$\hat{\kappa} \to \infty$ and $\xi \to 0$ we obtain  the classical
(undeformed) Poincare group ${\mathcal P}$. For ${\kappa} \to
\infty$ and $\xi \to 0$ we get the Lie-algebraically  twisted
classical Poincare group \cite{lukwor}, while in the case ${\kappa}
\to \infty$ and $\hat{\kappa} \to \infty$ we obtain the canonical
deformation of
 classical Poincare Hopf algebra ${\mathcal P}_\xi$ (see  \cite{3e}).

\section{{{Contractions to twisted ${\kappa}$-Galilei algebras and ${\kappa}$-Galilei groups}}}

In this section we calculate the  nonrelativistic contractions of
Hopf structures derived in previous sections, i.e. we find their
nonrelativistic counterparts - the canonical and Lie-algebraic twist
deformations of $\kappa$-Galilei algebra.


\subsection{{{Canonical deformation of ${\kappa}$-Galilei algebra}}}

Let us introduce the following standard redefinition of Poincar\'{e}
generators \cite{inonu} (see also \cite{barut})
\begin{equation}
P_{0 } = \frac{{\Pi_{0 }}}{c}\;\;\;,\;\;\;P_{i } =
\Pi_{i}\;\;\;,\;\;\;M_{ij}= K_{ij}\;\;\;,\;\;\;M_{i0}= cV_i\;,
\label{contr2}
\end{equation}
where parameter $c$ describes the light velocity. We start with
canonical twisted algebra $\,{\mathcal U}_{{\xi,\kappa}}({\mathcal
P})$, i.e. we introduce two  parameters $\overline{\kappa}$ and
$\overline{\xi}$ such that ${\kappa}=\overline{\kappa}/c$ and
${\xi}=\overline{\xi}c$. Next, one performs the contraction limit of
 algebraic part (\ref{a1})-(\ref{a5}) and co-sector
(\ref{modcopro1})-(\ref{modcopro2}) in two steps (see e.g.
\cite{con1}). Firstly, we rewrite the formulas (\ref{a1})-(\ref{a5})
and (\ref{modcopro1})-(\ref{modcopro2}) in term of the operators
(\ref{contr2}) and parameters $\overline{\kappa}$, $\overline{\xi}$.
Secondly, we take the $c\to \infty$ limit, and in such a way, we get
the following algebraic
\begin{eqnarray}
&&\left[\, K^{ij},K^{kl}\,\right] =i\left( \delta
^{il}\,K^{jk}-\delta
^{jl}\,K^{ik}+\delta^{jk}K^{il}-\delta ^{ik}K^{jl}\right) \;, \label{nnn1}  \\
&~~&  \cr &&\left[\, K^{ij},V^{k}\,\right] =i\left( \delta
^{jk}\,V^i-\delta ^{ik}\,V^j\right) \;\;\;,\;\;\;\left[
\,K^{ij},\Pi_{k }\right] =i\left( \delta ^{j}_{~ k }\,\Pi_{i
}-\delta ^{i}_{~ k }\,\Pi_{j }\right)\;, \label{nnn}
\\
&~~&  \cr &&~~~~~~~~~~~~\left[ \,V_i,V_j\,\right] =
0\;\;\;,\;\;\;\left[ \,V^i,\Pi_{0 }\,\right]
=i\Pi_i\;\;\;,\;\;\;\left[ \,\Pi_{\rho
},\Pi_{\sigma }\,\right] = 0\;\;\;,\;\;\;\\
&~~& \nonumber\\
&&~~~~~~~~~~\left[ \,V^i,\Pi_{j }\,\right] =
\delta^{i}_{~j}\frac{1}{2\overline{\kappa}}\overrightarrow{\Pi}^2
-\frac{1}{\overline{\kappa}}\Pi_{i }\Pi_{j
}\;\;\;,\;\;\;C_{\overline{\kappa}} = \overrightarrow{\Pi}^2{\rm
e}^{\frac{\Pi_0}{\kappa}}\;,\label{nnn2}
\end{eqnarray}\\
and coalgebraic
\begin{eqnarray}
\Delta_{{\overline{\xi},\overline{\kappa}}}(\Pi_0) &=&
 \Pi_0\otimes 1 + 1\otimes \Pi_0\;\;\;,\;\;\;
  \Delta_{{\overline{\xi},\overline{\kappa}}}(\Pi_i)  =
\Pi_i\otimes {\rm e}^{-\frac{\Pi_0}{\overline{\kappa}}} + 1\otimes \Pi_i\;,\label{wmodcopro1}\\
&~~&  \cr \Delta_{{\overline{\xi},\overline{\kappa}}}(K^{ij}) &=&
\Delta_{\overline{\kappa}}(K^{ij}) + \overline{\kappa}
\frac{\overline{\xi}}{2} \left( \delta^j_{~3}\Pi_i -
\delta^i_{~3}\Pi_j\right)
\otimes \left({\rm e}^{-\frac{\Pi_0}{\overline{\kappa}}} -1 \right)\;,\\
 &~~&  \cr
\Delta_{{\overline{\xi},\overline{\kappa}}}(V^{i}) &=&
\Delta_{\overline{\kappa}}(V^{i}) - \frac{\overline{\xi}}{2} \Pi_3
\otimes \Pi_i{\rm e}^{-\frac{\Pi_0}{\overline{\kappa}}} +\\
&~~&  \cr &+&{\frac{\overline{\xi}}{2}}\left( \delta^{i}_{~3}
\frac{1}{2\overline{\kappa}}\vec{\Pi}^2
 + \frac{1}{\overline{\kappa}}\Pi_i\Pi_3 \right) \otimes \overline{\kappa}\left({\rm
e}^{-\frac{\Pi_0}{\overline{\kappa}}} -1 \right){\rm
e}^{-\frac{\Pi_0}{\overline{\kappa}}}\\
&~~&  \cr
 &+& {\frac{\overline{\xi}}{2}}\left( \delta^j_{~3}\Pi_i - \delta^i_{~3}\Pi_j\right)
\otimes \Pi_j\left({\rm e}^{-\frac{\Pi_0}{\overline{\kappa}}} -1
\right)\;, \label{wmodcopro2}
\end{eqnarray}
sectors, where $\Delta_{\overline{\kappa}}(a) =
\Delta_{{\kappa}}(a)$. The antipodes look as follows
\begin{eqnarray}
S_{{\overline{\xi},\overline{\kappa}}}(\Pi_0) &=&
S_{\overline{\kappa}}(\Pi_0) = -\Pi_0\;\;\;,\;\;\;
S_{{\overline{\xi},\overline{\kappa}}}(\Pi_i) =
S_{\overline{\kappa}}(\Pi_i) = -\Pi_i{\rm
e}^{\frac{\Pi_0}{\overline{\kappa}}}\;,
~~~~~~~~~~~~~~\label{ddmodanti1}\\
&~~&  \cr ~~~~~~~~~~S_{{\overline{\xi},\overline{\kappa}}}(K^{ij})
&=& S_{\overline{\kappa}}(K^{ij})- \overline{\kappa}\overline{\xi}
\left(\delta^{j}_{~3}\Pi_i -
\delta^i_{~3}\Pi_j\right)\cdot\left(\exp({\Pi_0}/{\overline{\kappa}})-1
\right) \;,
\\
&~~&  \cr S_{{\overline{\xi},\overline{\kappa}}}(V^{i}) &=&
S_{\overline{\kappa}}(V^{i}) -
\overline{\xi}\left(\delta^{j}_{~3}\Pi_i - \delta^i_{~3}\Pi_j\right)
\Pi_j\cdot{\rm e}^{\frac{\Pi_0}{\overline{\kappa}}} \cdot
\\ &~~&  \cr &\cdot& \left(\exp({\Pi_0}/{\overline{\kappa}})-1 \right)-\overline{\xi}
\Pi_3\Pi_i 
{\rm
e}^{\frac{2\Pi_0}{\overline{\kappa}}}+\\
&~~&  \cr &-&\overline{\kappa}\overline{\xi}\left( \delta^i_{~3}
\frac{1}{2\overline{\kappa}}\vec{\Pi}^2  -
\frac{1}{\overline{\kappa}} \Pi_i\Pi_3\right)\cdot{\rm
e}^{\frac{\Pi_0}{\overline{\kappa}}}\cdot
\left(\exp({\Pi_0}/{\overline{\kappa}})-1 \right) \;,
\label{ddmodanti2}
\end{eqnarray}
with $S_{\overline{\kappa}}(a) = S_{{\kappa}}(a)$. The relations
(\ref{nnn1})-(\ref{ddmodanti2}) define the canonically twisted
$\kappa$-Galilei algebra $\,{\mathcal
U}_{\overline{\xi},\overline{\kappa}}({\mathcal G})$. One can see
that for $\overline{\xi} \to 0$ we get the
$\overline{\kappa}$-deformed Galilei group $\,{\mathcal
U}_{\overline{\kappa}}({\mathcal G})$ firstly studied in
\cite{{gal1}} (see also \cite{{con1}}). In   $\overline{\kappa} \to
\infty$ limit we obtain the canonically deformed 
algebra $\,{\mathcal U}_{\overline{\xi}}({\mathcal G})$  found in
\cite{{con3}}. Obviously, for $\overline{\kappa} \to \infty$ and
$\overline{\xi} \to 0$ one gets the undeformed Galilei quantum group
$\,{\mathcal U}_{0}({\mathcal G})$.

\subsection{{{Lie-algebraic deformation of ${\kappa}$-Galilei algebra}}}

In the case of Lie-algebraic modification of $\kappa$-Poincare
algebra, we perform contraction with respect the parameters
${\kappa}=\overline{\kappa}/c$ and ${\hat \kappa}=\overline{{\hat
\kappa}}/c$. Due to the relations
(\ref{fmodcopro1})-(\ref{fmodcopro2}) we obtain the coproducts
$\Delta_{{\overline{\kappa},\overline{\hat\kappa}}}(\Pi_\rho)$,
$\Delta_{{\overline{\kappa},\overline{\hat\kappa}}}(K^{ij})$ and
$\Delta_{{\overline{\kappa},\overline{\hat\kappa}}}(V^i)$ such that
$\Delta_{{\overline{\kappa},\overline{\hat\kappa}}}(a)
=\Delta_{{\cal F}_{{{\hat\kappa},{\kappa}}}}(a)$. In this way we get
the Lie-twisted Galilei algebra $\,{\mathcal U}_{\overline{\kappa},
\overline{\hat\kappa}}({\mathcal G})$, which for
$\overline{\hat\kappa} \to \infty$ passes into
$\overline{\kappa}$-deformed Galilei group $\,{\mathcal
U}_{\overline{\kappa}}({\mathcal G})$.

\subsection{{{Canonical and Lie-algebraic deformation of ${\kappa}$-Galilei group}}}

Finally, let us find the contraction of
($\hat{\kappa}$,$\xi$)-deformed Poincare  group ${\mathcal
P}_{\kappa,\hat{\kappa},\xi}$ (see (\ref{gr1})-(\ref{gr2}) and
(\ref{gkappa5}), (\ref{gkappa6})). In this purpose we introduce the
following redefinition of $\Lambda_{\ \nu}^{\mu}$, $a^\mu$
generators \cite{gg}
\begin{eqnarray}
&& {\Lambda} _{\ 0 }^{0 } = \left(
1+\frac{\overline{v}^2}{c^2}\right)^{\frac{1}{2}}\;\;\;,\;\;\;{\Lambda}
_{\ 0 }^{i } = \frac{v^i}{c}
\;\;\;,\;\;\; {\Lambda} _{\ i }^{0 }= \frac{v^k{R} _{\ i}^{k }}{c}\;,\label{zadrugizm1}\\
&~~&  \cr && {\Lambda} _{\ i }^{k } = \left(\delta _{\ l }^{k
}+\left(\left(
1+\frac{\overline{v}^2}{c^2}\right)^{\frac{1}{2}}-1\right)
\frac{v^kv^l}{\overline{v}^2}\right){R} _{\ i }^{l } \;,\\
&~~&  \cr &&a^i = b^i\;\;\;,\;\;\;a^0 = c\tau\;, \label{zadrugizm7}
\end{eqnarray}
where  $\{\,{R} _{\ j }^{i },v^i,\tau,b^i\,\}$ denote the generators
of  Galilei group. With use of the formulas
(\ref{zadrugizm1})-(\ref{zadrugizm7}) in the contraction limit $c
\to \infty$  we get
\begin{eqnarray}
&&[\,R^{k}_{~l},b^{i}\,] = -\frac{i}{\overline{\kappa}} (v^k
R^{i}_{~l} - \delta^{ki}v^{\rho}R^{\rho}_{~l})+
\frac{1}{{\overline{\hat\kappa}}}
v^i( \delta_{2l}R^{k}_{~1} - \delta_{1l}R^{k}_{~2})\;, \label{gagr1}  \\
&&~~\vspace{0.3cm} \nonumber \\
&&[\,R^{k}_{~l},\tau\,] =  \frac{1}{{\overline{\hat\kappa}}} ((
\delta_{2l}R^{k}_{~1} - \delta_{1l}R^{k}_{~2}) -
(\delta^{k}_{~2}R_{1l}
-\delta^{k}_{~1}R_{2l} ))\;,   \\
&&~~\vspace{0.3cm} \nonumber \\
&&[\,v^i,b^{j}\,] = -\frac{i}{\overline{\kappa}} (v^iv^j -
\frac{1}{2}\delta^{ij}\overline{v}^2)\;\;\;,\;\;\; [\,v^i,\tau\,] =
-\frac{i}{\overline{\kappa}} v^i-\frac{1}{{\overline{\hat\kappa}}}
(\delta^{i}_{~2}v_{1} -\delta^{i}_{~1}v_2 ) \;,\\
&&~~\vspace{0.3cm} \nonumber \\
&&[\,\tau,b^{i}\,] = -\frac{i}{\overline{\kappa}} b^{i}
+\frac{i}{\overline{{\hat\kappa}}}(\delta^{i}_{~2}b^{1}
-\delta^{i}_{~1}b^{2} ) +i\frac{\overline{\xi}}{2}(R^{i}_{~3}
+\delta^{i}_{~3})\;, \\
&&~~\vspace{0.3cm} \nonumber \\
&&[\; b^i,b^j\;]= i\frac{\overline{\xi}}{2}( v^{i }\,{R} _{\ 3 }^{j
}-{R} _{\
3 }^{i}v^j )\;,\;\;\; \\
&&~~\vspace{0.3cm}\nonumber\\ && [\; {R} _{\ j }^{i },{R} _{\ l
}^{k}\;]=[\; {v}^i,{R} _{\ l }^{k }\;]  = [\; v^i,v^j\;]
=0\;.\label{gagr2}
\end{eqnarray}
The coproducts remain undeformed
\begin{eqnarray}
&&\Delta ({R}_{\ j }^{i}) = {R}_{\ k }^{i }\otimes {R}_{\ j}^{k
}\;\;\;,\;\;\; \Delta ({v}^{i })={R}_{\ j}^{i}\otimes {v}^{j
}+{v}^{i
}\otimes 1\;,\label{toporzelxx}\\
&~~&  \cr &&\Delta (\tau)= \tau\otimes 1+1\otimes
\tau\;\;\;,\;\;\;\Delta ({b}^{i })={R}_{\ j}^{i }\otimes {b}^{j
}+{v}^{i }\otimes \tau+{b}^{i }\otimes 1\;. \label{toporzelxx1}
\end{eqnarray}
The relations (\ref{gagr1})-(\ref{toporzelxx1}) with classical
antipodes and counits define the ($\hat{\kappa}$,$\xi$)-deformed
Galilei group ${\mathcal G}_{\kappa,\hat{\kappa},\xi}$. As in the
case of relativistic symmetries,  for $\hat{\kappa}=\infty$ we get
 dual group to the Galilei algebra $\,{\mathcal
U}_{\overline{\xi},\overline{\kappa}}({\mathcal G})$, while for $\xi
=0$ we obtain dual partner for the   algebra $\,{\mathcal
U}_{\overline{\kappa}, \overline{\hat\kappa}}({\mathcal G})$.

Finally, one should also notice that in the
$\overline{{\hat\kappa}}\to \infty$ and $\overline{\xi} \to 0$
limits we get the well-known $\overline{{\kappa}}$-deformed Galilei
group $\,{\mathcal G}_{{\overline{{\kappa}}}}$ (see \cite{gg}),
while for $\overline{{\kappa}}\to \infty$ and $\overline{\xi} \to 0$
or $\overline{{\hat\kappa}}\to \infty$, we obtain the quantum
Galilei groups recovered in \cite{trojka}.

\section{{{Final remarks}}}

In this article we introduced two twist extensions of
$\kappa$-Minkowski spaces corresponding to   soft and Lie-algebraic
type of noncommutativity (see (\ref{dst2}) and (\ref{dst2ff})). For
such modified space-times we find  their quantum Poincare algebras
and corresponding dual quantum groups. The nonrelativistic
contractions are performed as well.

As it was mentioned in Introduction the Lie-algebraic  twist
introduces in natural way a second  mass-like parameter of
deformation. Consequently, in such a way, one can obtain a
"modification" of so-called Doubly Special Relativity
(\cite{dsr1a}-\cite{dsr1c})
 with one fundamental mass parameter, by introducing a second
observer-independent mass-like scale.

It should be also noted that  this paper is only a starting point
for a further investigation. For example, it is interesting to ask
about the noncommutative field theory given on such generalized
quantum Minkowski space-times. In particular, its formulation
requires the construction of a proper differential calculus, a
proper star product of fields, and a suitable deformation of
statistics for creation/annihilation operators (see e.g.
\cite{sitarz}-\cite{anglik}). The above problems are now under
considerations and they are postponed for further investigation.

\section*{Acknowledgments}
The author would like to thank  Jerzy Lukierski and  Mariusz
Woronowicz for many valuable discussions. There are also thanks for
Andrzej Borowiec, Jerzy Kowalski-Glikman and Marek Mozrzymas for
discussions on
classical r-matrices. \\
This paper has been financially supported by Polish funds for
scientific research (2008-10) in the framework of a research
project.

\end{document}